# Reading Copom's Tone: A Weighted LLM Framework for Hawkish-Dovish Sentiment, Forward Guidance, and Uncertainty

*Sample used in this version: 80 policy statements, August 2016 to August 2026*

*Gabriel de Macedo Santos*

*Instituto de Tecnologia e Liderança*

**Abstract.** This paper documents an applied natural-language-processing framework for measuring the tone of Brazilian Monetary Policy Committee (Copom) statements. The project is explicitly inspired by iSent, Itaú's Central Bank sentiment classifier, particularly its sentence-level division of official communication into hawkish, dovish, neutral, and out-of-context classes. The implementation extends that idea in three directions. First, an LLM identifies short hawkish and dovish expressions and assigns each a 0-to-1 intensity weight. Second, the document index combines sentence counts with document-specific average signal intensities, producing a bounded score from -1 to 1. Third, a separate full-document layer measures forward-guidance direction, guidance explicitness, uncertainty level, and change in uncertainty. The empirical sample is restricted to communications dated August 2016 or later and contains 80 statements and 1,498 classified sentences from August 31, 2016 through August 5, 2026. Across this sample, 33.3% of sentences are hawkish, 18.0% dovish, 42.1% neutral, and 6.5% out of context. The average document score is +0.107, while the most hawkish reading is +0.570 in August 2021. The latest statement, dated August 5, 2026, scores +0.232, with eight hawkish, two dovish, and nine neutral sentences. Its structural overlay is more nuanced: guidance is directionally ambiguous but partly explicit, while uncertainty is classified as central and higher than at the prior meeting. Tone and the guidance-direction score have a contemporaneous Pearson correlation of 0.719. These are descriptive outputs, not a validated forecast of Selic decisions or DI returns. The main contribution is therefore methodological: a transparent, incremental, auditable system that separates rhetorical tone from policy guidance and uncertainty.

## 1. Introduction

Central bank communication is not ancillary to monetary policy. It shapes expectations about the reaction function, the likely path of short-term interest rates, the balance of risks, and the conditions under which policymakers may change course. In inflation-targeting regimes, the language surrounding a rate decision can therefore carry information that is distinct from the mechanical decision itself. Blinder et al. (2008) describe communication as a potentially powerful part of the central bank's toolkit, while Hansen and McMahon (2016) show that communication about forward guidance can matter more for markets than descriptions of current economic conditions.

The practical difficulty is measurement. Copom statements contain simultaneous signals about inflation, activity, the exchange rate, fiscal policy, external risks, expectations, and the degree of confidence in the baseline scenario. A single statement may describe easing in current activity while warning about unanchored inflation expectations. A pure word count can miss this context, while a document-level LLM label can hide internal disagreement across sentences.

This project addresses that problem with a sentence-level, prompt-based classifier and a separate structural overlay. Its starting inspiration is iSent, introduced by Itaú Unibanco Macro Research in July 2024. iSent classifies sentences from official central-bank documents as dovish, neutral, hawkish, or out of context, and aggregates the relative presence of those classes into an index between -1 and 1. The current project adopts that intuitive unit of analysis and class taxonomy, while deliberately changing the calibration and output architecture.

The framework adds phrase-level intensity weights, treats formal decision-only sentences as neutral by rule, and estimates forward guidance and uncertainty independently of the tone score. It also uses an incremental cache so that new statements can be processed without re-annotating the historical corpus. These additions are useful because hawkish tone, explicit guidance toward a hike, and high uncertainty are related but not identical objects. The latest reading illustrates the distinction: the August 2026 statement is hawkish in tone, but its policy guidance is directionally ambiguous.

The paper makes three contributions. First, it documents the exact scoring formula and the data lineage from sentence annotations to document scores. Second, it presents a consistently defined historical sample from August 2016 to August 2026, excluding all earlier observations from the empirical analysis. Third, it provides an implementation audit that separates what is currently operational from what remains a validation or research extension. The result is best viewed as a monitoring index, not yet as a trading signal or causal estimate of communication effects.

## 2. Related Literature and the iSent Inspiration

The empirical literature on central-bank text has moved from dictionaries and bag-of-words representations toward contextual language models. Correa et al. (2021) construct a domain-specific dictionary for financial-stability reports, illustrating the value of tailoring sentiment vocabularies to policy language. Hansen and McMahon (2016) use computational linguistics to separate communication about economic conditions from forward guidance. More recent work uses models trained or adapted to the central-banking domain, including CentralBankRoBERTa (Pfeifer and Marohl, 2023) and the CB-LMs introduced by Gambacorta et al. (2024).

LLMs reduce some of the rigidity of dictionaries by interpreting context, negation, conditionals, and policy-specific phrasing. They also create new risks: model outputs can vary with prompts and versions; pretraining data may contain the documents being classified; and reproducibility depends on pinning the model and the inference configuration. Silva, Moriya, and Veyrune (2025) show the potential of sentence-level LLM classification at large scale, but their framework also underscores the importance of separating communication dimensions rather than collapsing all language into one sentiment label.

The closest direct inspiration is Itaú's iSent. Its published methodology uses roughly one thousand economist-labeled Brazilian sentences, sentence-level GPT-4 classification, and retrieved similar examples through a FAISS index. The iSent index is the difference between hawkish and dovish sentence counts divided by the number of hawkish, dovish, and neutral sentences. Itaú reports correlations of 0.79 with the contemporaneous Selic change and 0.77 with the one-meeting-ahead change for Brazil. Those figures are properties of iSent and should not be attributed to the index developed here.

| Dimension | iSent (Itaú, 2024) | This project |
|---|---|---|
| Unit of analysis | Sentence | Sentence for tone; full document for structural dimensions |
| Tone classes | Hawkish, dovish, neutral, out of context | Hawk, dove, neutral, out |
| Model grounding | Economist-labeled examples; GPT-4; FAISS retrieval | Prompt instructions and three heuristic anchors; API-hosted LLM |
| Index | Relative class presence; unweighted | Class counts scaled by document-specific mean phrase intensity |
| Additional dimensions | Published policy/market comparison and rates backtest | Guidance direction, explicitness, uncertainty level, uncertainty change |
| Retrieval in baseline | Yes | No; the current FAISS routine is optional and not called by sentence classification |
| Current validation status | Published correlation and backtest | Descriptive implementation; external validation still required |

***Table 1. Methodological relationship between iSent and the current project.***

*Note: The project is independently developed and is not affiliated with or endorsed by Itaú Unibanco. iSent is cited as methodological inspiration.*

## 3. Data and Project Architecture

The empirical sample covers 80 Copom statement dates from August 31, 2016 through August 5, 2026. The August 2016 cutoff is applied before computing any descriptive statistic, sentence share, correlation, extreme reading, or regime average in this paper. The primary source is the official Portuguese text of Copom statements. The code expects one record per date, removes rows with missing dates or text, sorts chronologically, and retains the last observation when duplicate dates exist. The supplied analytical files contain 1,498 classified sentence objects within the retained window. Of those, 1,400 enter the score denominator; 98 out-of-context sentences are excluded.

The raw statements file and external model configuration are not included in the supplied artifact set. The annotations, scores, structural outputs, and code are sufficient to reproduce the descriptive results in this paper, but strict end-to-end replication would also require the raw statements, model identifier, prompt version, and inference settings. This distinction is recorded in the implementation audit rather than obscured.

| Artifact | Observed content | Role in the framework |
|---|---|---|
| sentiment_classifier.py | 613-line incremental pipeline | Sentence splitting, LLM prompts, scoring, structural classification, plot, optional FAISS |
| sentence_annotations.json | 80 retained dates; sentence labels and phrase signals | Primary audit trail after the August 2016 sample filter |
| scores_comunicados.json | 80 retained document scores and class counts | Weighted tone index for the study window |
| structural_analysis.json | 80 retained document-level analyses | Guidance direction/explicitness and uncertainty level/change |
| pesos_hawk_dove.json | Full-file diagnostic weights | Not used in sample statistics; scores use document-specific means |
| text_features_clean.csv | 78 statements, Aug. 2016-Apr. 2026 | Convenience table; JSON adds June and August 2026 |
| hawk_dove_history.png | History plot from Aug. 2016 | Published visualization supplied with the project |

***Table 2. Uploaded artifacts and their economic or implementation role.***

*Source: Uploaded project files. Counts were recomputed from the supplied JSON and CSV outputs.*

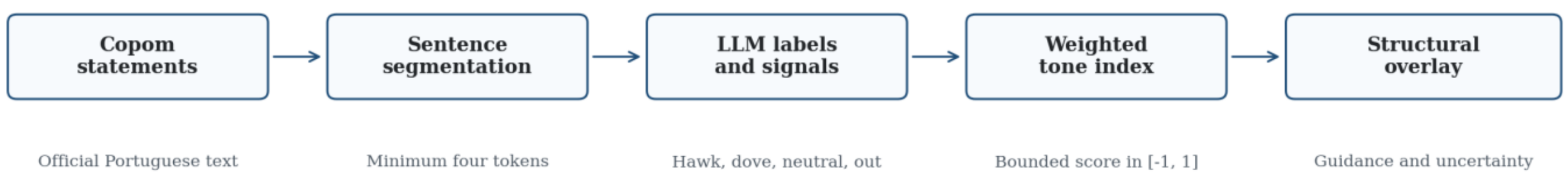

***Figure 1. Project pipeline from official text to the tone index and structural overlay.***

*Source: Author's implementation based on the uploaded classifier.*

## 4. Methodology

### 4.1. Text normalization and sentence segmentation

The pipeline first converts the statements into a two-column date-text dataset. Dates are parsed, invalid observations are dropped, and duplicate dates are deduplicated. New lines are replaced with spaces before classification. Sentence boundaries are identified with a regular expression that looks for whitespace following periods, question marks, exclamation points, semicolons, or colons when the next token begins with an uppercase letter, number, quotation mark, or parenthesis. Segments with fewer than four whitespace-delimited tokens are excluded.

This deterministic splitter is efficient, but it inherits formatting defects from scraped source text. Several archived annotations contain missing spaces after punctuation, causing logically separate sentences to be merged. This matters because a merged segment can contain both hawkish and dovish phrases while receiving only one final class. The structural full-document layer is less exposed to this segmentation issue, but the tone score is not.

### 4.2. Sentence classification and phrase-level signals

Each segment is sent to an LLM under a Portuguese prompt that defines four mutually exclusive classes. Hawkish sentences emphasize persistent or elevated inflation, upside risks, restrictive policy, higher rates, fiscal deterioration, or other forces that increase the required degree of monetary restraint. Dovish sentences emphasize disinflation, economic slack, weak activity, easing, or prospective rate cuts. Neutral sentences are informational or balanced. Out-of-context sentences include administrative content, names, and material outside monetary policy or macroeconomic conditions.

The prompt also asks the model to extract short original expressions that support a hawkish or dovish reading. Each signal is assigned a weight from 0 to 1: 0.1-0.3 for weak, 0.4-0.6 for moderate, and 0.7-1.0 for strong evidence. Weights and labels are validated and clipped after parsing. The model can return signals of both signs inside one segment, even though the segment has one final class. Up to forty segments are classified per API call.

A special instruction treats a sentence that only reports the mechanical rate decision as neutral with no signals. This is economically motivated: the index is intended to capture the communication surrounding the action, not mechanically encode the rate change twice. The prompt further provides special guidance for the balance of inflation risks and includes three historical calibration anchors: March 2020 as strongly dovish, May 2021 as hawkish, and August 2023 as dovish. These anchors guide the LLM; they are not hard constraints on the final score, which is recomputed mechanically from the returned labels and weights.

### 4.3. Weighted document score

For statement t, let $N_{H,t}$, $N_{D_t}$, and $N_N$ denote the numbers of hawkish, dovish, and neutral sentences. Out-of-context sentences do not enter the denominator. Let the average intensity of all hawkish signals extracted anywhere in the document be $w|H_t$ , and define $w|D_t$ analogously. The implementation uses a default intensity of 1 when no signal of a given sign is available. The raw score is:

$$Score_t = \frac{w|H_t \times N_{H,t} - w|D_t \times N_{D,t}}{N_H + N_D + N_N}$$

The score is then clipped to the interval [-1, 1]. Positive values indicate a net hawkish tone and negative values a net dovish tone. Neutral sentences dilute the magnitude without changing the numerator. A subtle but important implementation detail is that intensity averages are computed from all extracted signals, not only from signals found in sentences whose final class matches the signal. Thus, a neutral or hawkish segment that contains a dovish phrase can affect the dovish intensity term.

Within the retained August 2016-August 2026 sample, hawkish and dovish signal weights average 0.654 and 0.586. These values are diagnostics and are not used by compute_scores. The live document score instead recalculates both intensities within each statement. This makes the index sensitive to local rhetorical strength but also increases sampling noise in short documents.

### 4.4. Structural communication layer

A second LLM call evaluates the entire statement along four dimensions. Guidance direction equals -1 when the text points to easing, 0 when the next move is unclear, and +1 when it points to tightening. Guidance explicitness equals 0 for no guidance, 0.5 for indirect or conditional guidance, and 1 for explicit guidance. Uncertainty level ranges from 0 to 3, while uncertainty change takes values -1, 0, or +1 relative to the prior meeting.

$$GuidanceScore_t = \text{Direction}_t \text{ x } \text{Explicitness}_t$$

The product preserves direction while shrinking indirect signals toward zero. It is not part of the tone index. This separation allows the framework to identify statements that are hawkish in diagnosis but neutral about the next policy action, or dovish in guidance while emphasizing unusually high uncertainty.

### 4.5. Incremental execution and optional retrieval

The pipeline is incremental. Existing annotations and structural classifications are loaded from JSON caches, and only dates not already present are sent to the LLM. The complete archive is then rescored so that all outputs remain synchronized. A force flag discards the cache and re-annotates the full sample. This architecture reduces cost and latency, but cached results also mean that historical classifications can reflect different model versions unless model and prompt metadata are stored with every run.

The code can optionally build a FAISS vector index over full statement texts. In the supplied implementation, however, the sentence-classification function does not query that index or inject retrieved examples into the prompt. The baseline should therefore be described as prompt-based LLM classification with optional retrieval infrastructure, not as an operational RAG classifier.

## 5. Results

### 5.1. Corpus composition and signal intensities

The August 2016-August 2026 sample contains 1,498 classified segments. Neutral sentences are the largest class, at 631 observations or 42.1% of the corpus. Hawkish sentences total 499 (33.3%), dovish sentences 270 (18.0%), and out-of-context sentences 98 (6.5%). The prevalence of neutral language is economically plausible because Copom statements contain projections, descriptions, decision mechanics, and administrative material that should not all be mapped into a policy stance.

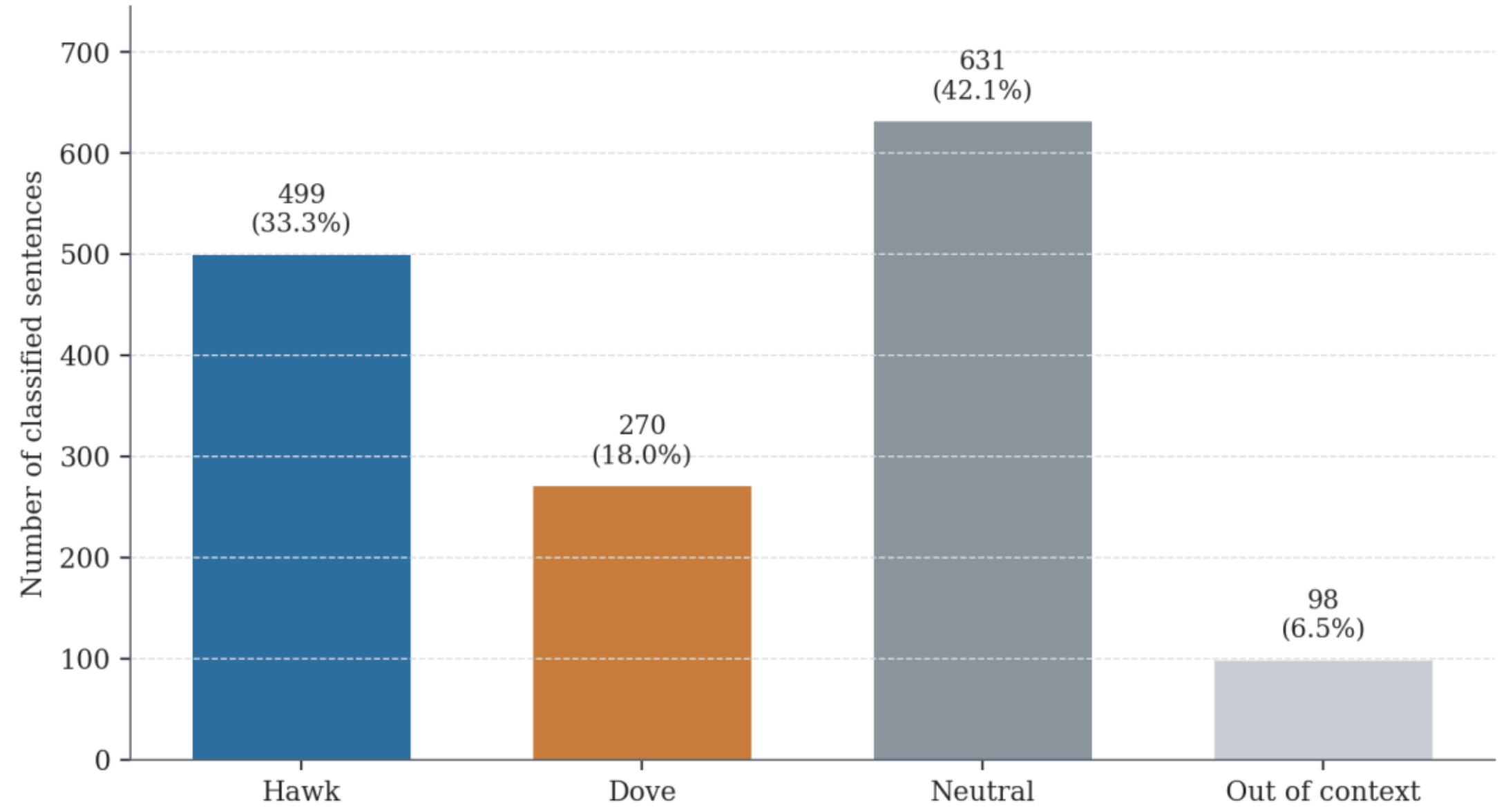

***Figure 2. Sentence-level class composition, August 2016-August 2026.***

*Note: Out-of-context sentences are excluded from the document-score denominator.*

The LLM extracted 1,552 hawkish phrase signals and 787 dovish signals within the retained sample. Their mean weights are 0.654 and 0.586, respectively, with standard deviations close to 0.20. Signal counts exceed sentence counts because one segment can contain multiple relevant phrases. The difference in average intensity contributes to a structural positive tilt in the numerator, but the actual score uses document-specific rather than global averages.

| **Metric** | **Aug. 2016-Aug. 2026 value** | **Interpretation** |
|---|---|---|
| Statements | 80 | August 31, 2016-August 5, 2026 |
| Classified segments | 1,498 | Includes out-of-context segments |
| Segments in denominator | 1,400 | Hawk + dove + neutral |

| Metric | Aug. 2016-Aug. 2026 value | Interpretation |
|---|---|---|
| Mean document score | +0.107 | Positive/hawkish tilt on average |
| Median document score | +0.106 | Typical statement is mildly hawkish |
| Standard deviation | 0.208 | Substantial meeting-to-meeting variation |
| Interquartile range | -0.041 to +0.265 | Middle 50% of document scores |
| Negative / nonnegative scores | 26 / 54 | Sign classification used by the history plot |

***Table 3. Descriptive statistics of the weighted Copom tone index.***

### 5.2. Historical path and monetary-policy regimes

The retained history shows a clear cyclical pattern. From August 2016 through 2020, communication is dovish on average (-0.078), consistent with repeated disinflation language, slack, and easing guidance. The series turns sharply positive in 2021-2023, with an average of +0.263, as inflation persistence, risk asymmetry, and the need for above-neutral policy dominate the communication. The 2024-August 2026 window remains strongly positive at +0.237, even though its average guidance score is near zero. That combination reflects a communication regime in which the diagnosis can remain restrictive while the immediate direction of policy becomes conditional.

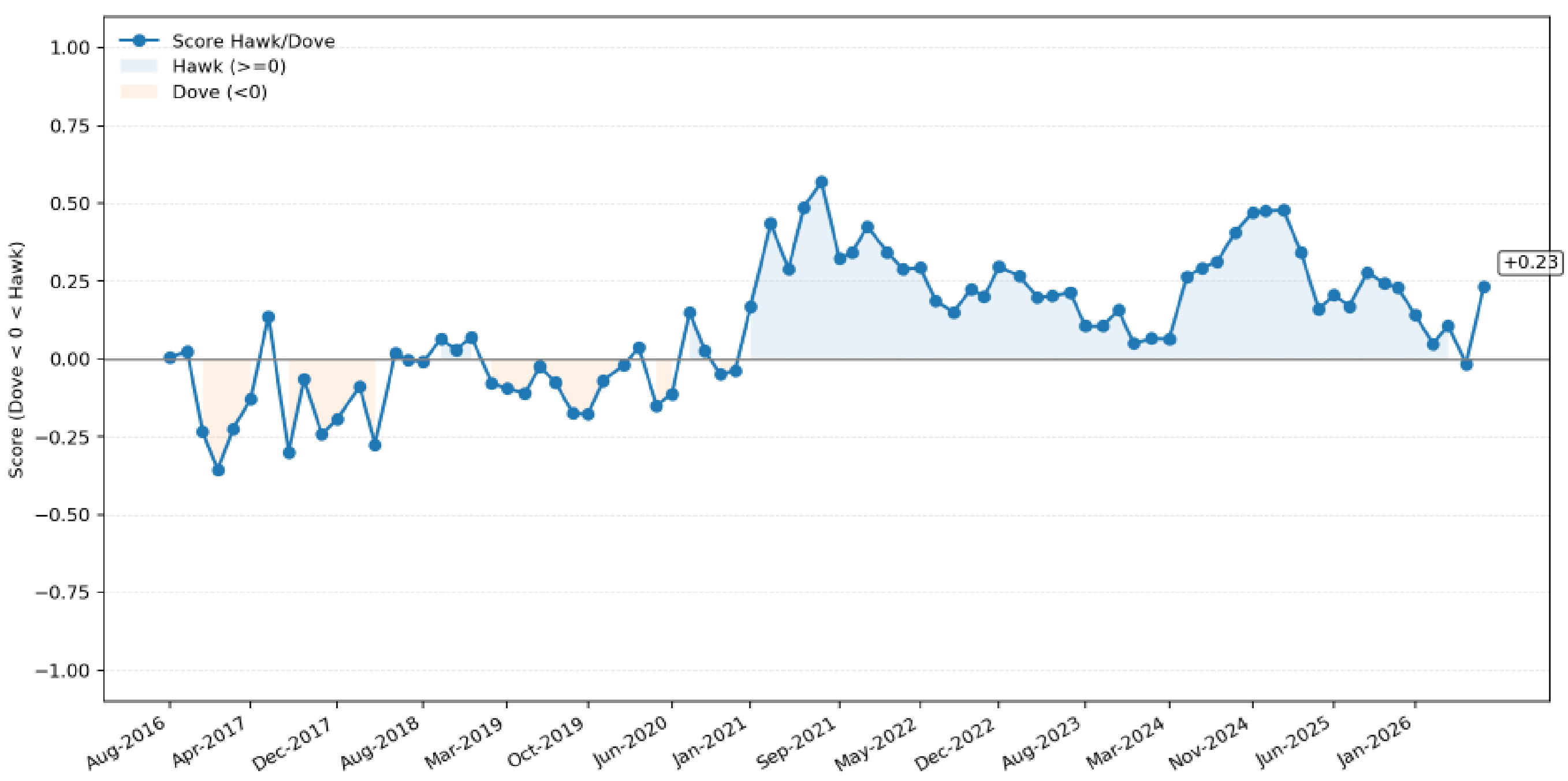

***Figure 3. Weighted hawkish-dovish score for the 80-statement study sample.***

*Note: Positive readings are hawkish; negative readings are dovish. All observations before August 2016.*

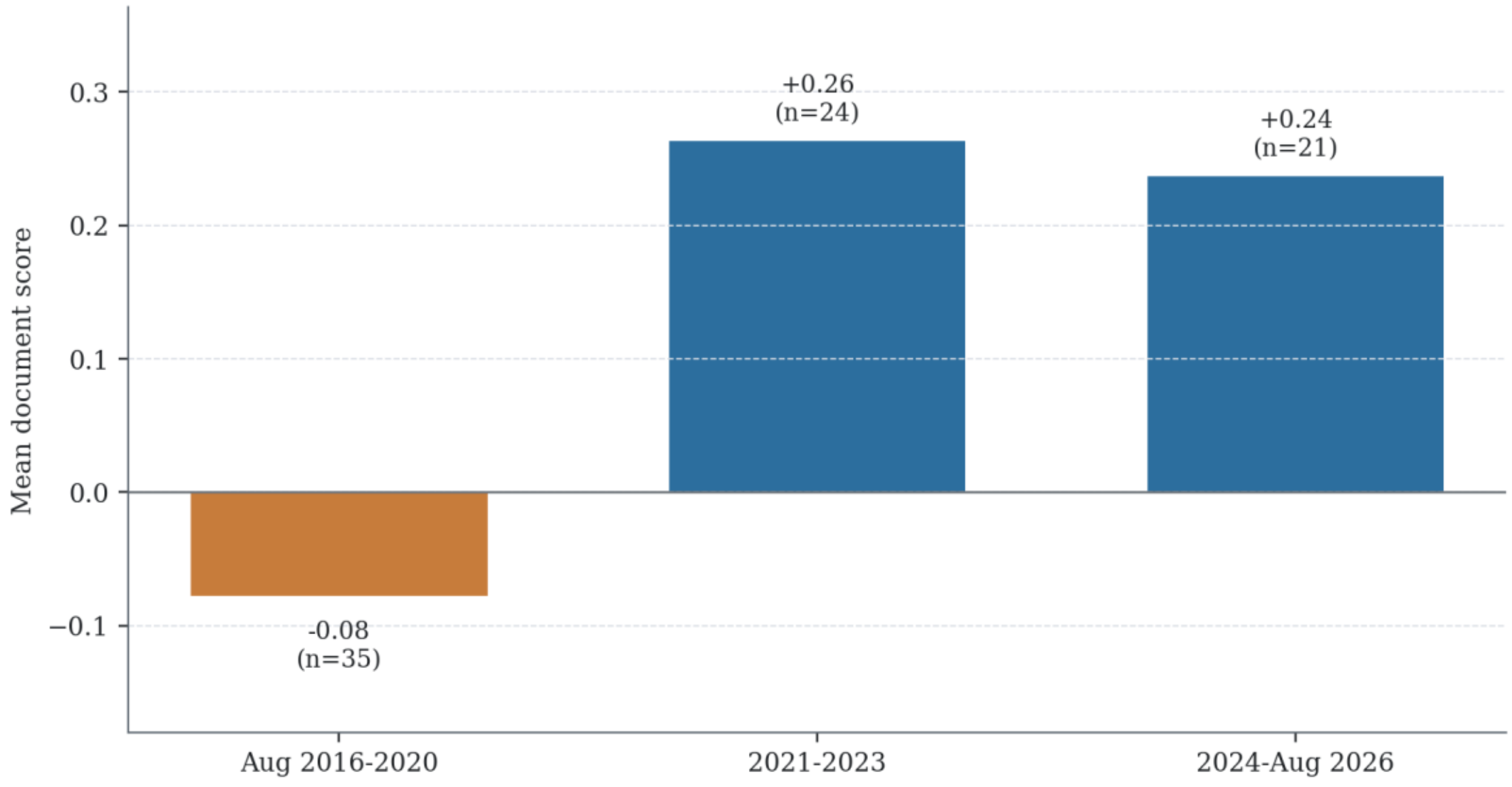

***Figure 4. Mean document score by broad communication regime.***

*Note: Regime boundaries are descriptive groupings selected for this paper, not estimated breakpoints.*

| Period | Documents | Mean tone | Mean guidance | Mean uncertainty |
|---|---|---|---|---|
| Aug. 2016-2020 | 35 | -0.078 | -0.457 | 2.23 |
| 2021-2023 | 24 | +0.263 | +0.458 | 2.54 |
| 2024-Aug. 2026 | 21 | +0.237 | -0.024 | 2.76 |

***Table 4. Era-level averages of tone, guidance, and uncertainty.***

*Note: Guidance equals direction multiplied by explicitness. Uncertainty ranges from 0 to 3.*

## 5.3. Extreme readings

The most dovish observation in the retained sample is January 11, 2017, at -0.357, with two hawkish, nine dovish, and two neutral sentences. The reading reflects an explicit intensification of the easing cycle amid weak activity and ongoing disinflation. The strongest hawkish reading is August 4, 2021, at +0.570, supported by fourteen hawkish sentences and no dovish sentences. That statement emphasizes persistent inflation, an unfavorable composition of prices, elevated fiscal risk, an upside-skewed balance of risks, and an explicit expectation of another adjustment of the same magnitude.

The largest positive readings cluster around the 2021 tightening cycle and the renewed hawkish communication of late 2024 and early 2025. The largest negative readings cluster around the 2016-2018 easing cycle, when disinflation, economic slack, and forward guidance toward continued cuts were recurrent. These patterns are economically coherent, but coherence is not the same as out-of-sample predictive validation.

| Date | Score | Hawk | Dove | Neutral | Reading |
|---|---|---|---|---|---|
| 2017-01-11 | -0.357 | 2 | 9 | 2 | Most dovish; easing intensification |
| 2017-07-26 | -0.300 | 0 | 7 | 6 | Broadly dovish easing language |
| 2021-08-04 | +0.570 | 14 | 0 | 4 | Most hawkish |
| 2021-06-16 | +0.487 | 13 | 1 | 3 | Tightening cycle |
| 2025-01-29 | +0.479 | 11 | 0 | 6 | Persistent inflation risks |
| 2026-08-05 | +0.232 | 8 | 2 | 9 | Latest reading |

***Table 5. Selected extreme and recent document scores.***

*Source: Author's calculations from the uploaded score and annotation files.*

## 5.4. Tone, forward guidance, and uncertainty

Guidance direction is dovish for 32 statements, neutral or ambiguous for 22, and hawkish for 26. Guidance is fully explicit in 44 statements and partial or conditional in 36; none of the retained statements is classified as having no guidance. Uncertainty is relevant or central (levels 2 or 3) in all 80 statements. The uncertainty-change variable is mostly zero: only 13 increases and three decreases are recorded, which partly reflects the prompt instruction to use zero when comparison with the prior meeting is not supported.

The contemporaneous Pearson correlation between the tone score and guidance direction multiplied by explicitness is 0.719; the Spearman correlation is 0.718. This is a strong but incomplete association. It confirms that tone and guidance often move together, while leaving substantial room for divergence. Because both outputs are generated by LLM prompts applied to the same documents, this correlation is descriptive and should not be interpreted as independent validation.

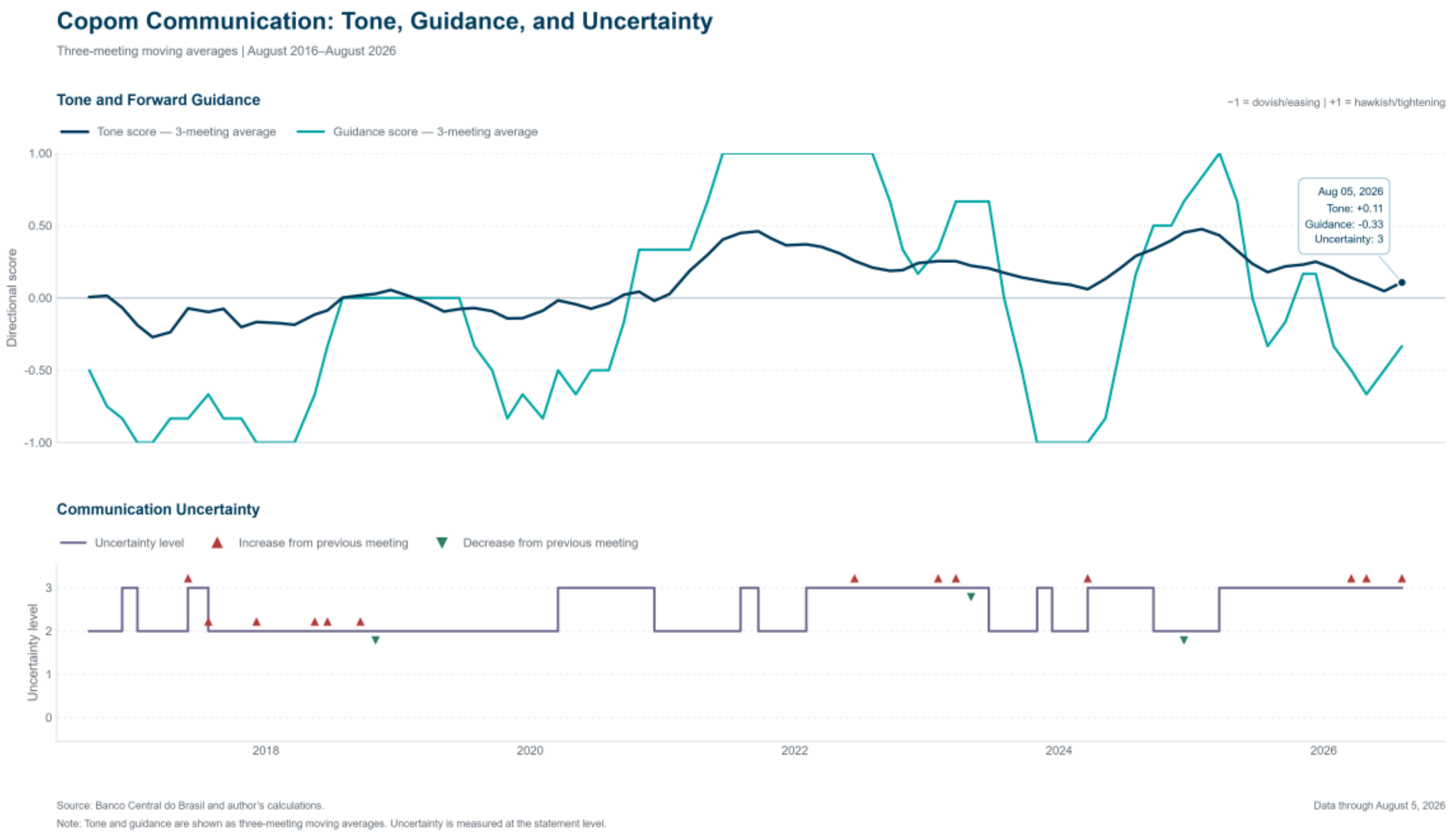

***Figure 5. Tone, forward guidance, and uncertainty across Copom statements.***

*Note: Tone and guidance use three-document moving averages for readability. Uncertainty is shown at the document level.*

## 5.5. Latest reading: August 5, 2026

The latest statement scores +0.232. The classifier identifies eight hawkish, two dovish, and nine neutral sentences. Hawkish content includes an upside-skewed balance of inflation risks, unanchored expectations, a resilient labor market, services inflation risk, exchange-rate depreciation risk, and the possibility that demand stimulus weakens monetary transmission. Dovish content includes the gradual moderation of activity, disinflation in headline and underlying measures, and downside scenarios tied to a sharper domestic or global slowdown.

The structural layer assigns guidance direction 0 and explicitness 0.5. In other words, the statement contains conditional guidance but does not clearly commit to either continued cuts or renewed tightening. Uncertainty is level 3 and change +1, reflecting the centrality of external conflict, asset and commodity volatility, fiscal concerns, and the statement's explicit description of a significant increase in uncertainty. The economic interpretation is therefore not simply 'hawkish.' It is a hawkish risk diagnosis combined with conditional, directionally ambiguous next-step guidance.

| Latest metric | August 5, 2026 value | Interpretation |
|---|---|---|

| Latest metric | August 5, 2026 value | Interpretation |
|---|---|---|
| Weighted tone score | +0.232 | Net hawkish diagnosis |
| Sentence mix | 8 hawk / 2 dove / 9 neutral | Hawkish content dominates directional sentences |
| Guidance direction | 0 | No clear next-move direction |
| Guidance explicitness | 0.5 | Conditional or partial guidance |
| Uncertainty level | 3 | Uncertainty is a central axis of the communication |
| Uncertainty change | +1 | Higher than at the prior meeting |

***Table 6. Latest Copom communication reading in the study sample.***

*Source: Uploaded scores, sentence annotations, and structural analysis.*

# 6. Validation, Robustness, and Limitations

The strongest feature of the project is auditability. Every score can be traced to sentence labels, extracted expressions, and weights. The code normalizes labels, clips weights, excludes out-of-context segments, and records structural justifications. The incremental cache makes the system operational for recurring Copom monitoring.

The principal limitation is the absence of a manually labeled holdout set. The same LLM that interprets the text supplies both the class and the signal weight. Without economist annotations, confusion matrices, calibration curves, or agreement statistics, the model's accuracy cannot be quantified. The three prompt anchors provide qualitative orientation, but their target values do not constrain the final mechanical score and the realized scores differ materially from the anchor suggestions.

A second limitation is segmentation. Missing spaces after punctuation can merge independent clauses or sentences. Because one merged segment receives one class, the output can understate mixed communication. A dedicated Portuguese sentence tokenizer, source-text cleaning, and unit tests around abbreviations, tables, and enumerated risk lists would reduce this problem.

A third limitation concerns intensity aggregation. The score multiplies a sentence count by the mean weight of every same-sign signal in the document. This is not equivalent to summing sentence-level weighted probabilities. It gives identical count weight to a strongly hawkish sentence and a weakly hawkish sentence while scaling the entire hawkish count by the document's mean signal intensity. Short documents and documents with many signals in one sentence can therefore behave differently from an intuitive additive scheme. An unweighted iSent-style index and a sentence-level weighted sum should be reported as robustness alternatives.

Finally, the project does not yet test whether the score predicts Selic changes, DI returns, yield-curve shifts, or forecast errors. Itaú's reported correlations and backtest belong to iSent. Replicating that validation here would require merging the score with the exact decision change, the one-meeting-ahead change, and properly lagged market prices, while preventing look-ahead bias. Trading results should include execution timing, contract construction, volatility scaling, transaction costs, and drawdown diagnostics.

| Robustness test | Purpose | Recommended output |
|---|---|---|
| Economist-labeled gold set | Measure classification validity | Precision, recall, F1, confusion matrix, Krippendorff alpha |
| Repeated LLM runs | Measure stochastic instability | Label agreement and score dispersion by statement |
| Alternative model/prompt | Assess model dependence | Rank correlation and extreme-reading stability |
| Unweighted index | Benchmark against iSent-style aggregation | Difference from weighted baseline |
| Sentence-level weighted sum | Test current mean-weight aggregation | Level and turning-point comparison |

| Robustness test | Purpose | Recommended output |
|---|---|---|
| Tokenizer ablation | Quantify segmentation sensitivity | Reclassified segments and score revisions |
| Selic lead/lag study | Test policy alignment | Contemporaneous and one-meeting-ahead correlations |
| DI event study/backtest | Test market relevance | Returns, information ratio, costs, drawdowns, OOS split |

***Table 7. Priority validation and robustness program.***

*Note: These tests are recommendations; their results are not present in the uploaded artifacts.*

# 7. Conclusion

This paper documents a weighted LLM framework for tracking the hawkish-dovish tone of Copom statements. Inspired by Itaú's iSent classifier, the project uses sentence-level hawk, dove, neutral, and out labels, but extends the architecture with phrase intensity weights and a separate analysis of forward guidance and uncertainty. The result is an incremental monitoring pipeline whose outputs can be traced back to individual sentences and supporting expressions.

Across 80 statements from August 2016 to August 2026, the average tone score is +0.107. Communication is dovish on average from August 2016 through 2020, sharply hawkish in 2021-2023, and remains positive in 2024-2026. The latest score is +0.232, but the structural layer shows why a single label is insufficient: forward guidance is conditional and directionally ambiguous, while uncertainty is central and rising.

The project's most credible current contribution is measurement and organization, not prediction. It provides a disciplined way to separate the tone of the economic diagnosis from the direction and explicitness of future policy guidance. Converting this monitoring system into a publishable predictive index requires a human-labeled benchmark, model-version controls, segmentation improvements, aggregation ablations, and a genuinely out-of-sample Selic and DI validation exercise.

# Appendix A. Implementation Audit

| Audit item | Observed implementation | Assessment |
|---|---|---|
| Date coverage | 80 matching dates after the August 2016 filter | Internally aligned study sample |
| Score denominator | Hawk + dove + neutral; out excluded | Consistent with published iSent-style denominator |
| Intensity calibration | Per-document signal means; global means saved separately | Article reports exact live logic |
| Decision-only rule | Prompt instructs neutral, but some historical outputs violate it | Requires post-processing or gold-set audit |
| Historical anchors | Three heuristic dates in the prompt | Guides LLM but does not constrain final score |
| Structural score | Direction x explicitness | Correctly separated from tone |
| FAISS | Index builder exists behind optional flag | Not queried in the classification path |
| Incremental cache | Only unseen dates are classified | Efficient, but requires model-version metadata |
| Raw corpus/config | Expected by code but absent from supplied archive | Blocks exact end-to-end replication |
| Feature CSV | 78 rows through April 2026 | JSON adds June and August 2026, yielding 80 study dates |

***Appendix Table A1. Implementation audit of the uploaded Copom sentiment project.***

# Appendix B. Scoring Interpretation

The index is a tone measure, not a probability. A score of +0.23 does not mean a 23% probability of a rate increase. It means that the weighted balance of hawkish and dovish sentence counts, after dilution by neutral content, equals +0.23 under the current aggregation rule. Likewise, zero can arise from genuinely balanced language, offsetting weighted counts, or a statement dominated by neutral segments.

Cross-period comparisons should consider document length and source formatting. Earlier statements within the retained sample can be shorter than recent statements, making their scores more discrete and more sensitive to a single classification. The score is therefore most reliable as one input to a broader communication dashboard that also reports sentence counts, guidance, uncertainty, and the underlying excerpts.